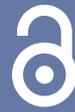

# CCOM-HuQin: An Annotated Multimodal Chinese Fiddle Performance Dataset

**DATASET**

YU ZHANG
ZIYA ZHOU
XIAOBING LI
FENG YU
MAOSONG SUN

*Author affiliations can be found in the back matter of this article

## ABSTRACT

HuQin (胡琴) is a family of traditional Chinese bowed string instruments. Playing techniques (PTs) embodied in various playing styles add abundant emotional coloring and aesthetic feelings to HuQin performance. The complex applied techniques make HuQin music a challenging source for fundamental MIR tasks such as pitch analysis, transcription and score-audio alignment. In this paper, we present a multimodal performance dataset of HuQin music that contains audio-visual recordings of 11,992 single PT clips and 57 annotated musical excerpts of classical pieces. We systematically describe the HuQin PT taxonomy based on musicological theory and practical use cases. Then we introduce the dataset creation methodology and highlight the annotation principles featuring PTs. We analyze the statistics in different aspects to demonstrate the variety of PTs played in HuQin subcategories and perform preliminary experiments to show the potential applications of the dataset in various MIR tasks and cross-cultural music studies. Finally, we propose future work to extend the dataset.

# 1. INTRODUCTION

As a Chinese traditional bowed string instrument family, HuQin(胡琴) has developed for over a thousand years since the Tang dynasty. As a significant component in Chinese music culture, HuQin music expresses the spirit of Chinese philosophy and reflects the social values in different historical contexts. HuQin originates from XiQin (奚琴, shown in Figure 1), first played by the Mongolic (northwestern China) Xi tribe. During the Yuan Dynasty, the name Xi became Hu (胡), representing ethnic minorities living in China's northern and western borderland. In the long course of history, HuQin became rooted in folk music and developed with traditional Chinese opera in local dialects. Nowadays, HuQin is a crucial leading and accompanying instrument group in the ensemble performance of Chinese folk music. Despite being in the same family, different HuQin instruments can be distinguished from each other by the tone quality and playing ranges that demonstrate unique characteristics of performance styles (Qiao et al., 2010).

The HuQin family encompasses various fiddles, of which we choose the most representative ones: Erhu (二胡), Banhu (板胡), Gaohu (高胡), Zhuihu (坠胡) and Zhonghu (中胡) (shown in Figure 1). Erhu has developed into different schools in different regions, such as Qin (秦, Northern) Erhu and Jiang-nan (江南, Southern) Erhu. The reform of Erhu's performance and composition in the 20th century makes it the currently best-known HuQin. Banhu, which has been popular in northern China for 300 years, mainly contains four subcategories: Soprano Banhu (SBanhu), Alto Banhu (ABanhu), Tenor Banhu (TBanhu) and Bass Banhu (BBanhu). The pitch ranges, tone quality and physical size of bound boxes distinguish them. Banhu is usually used as an accompaniment instrument in traditional local operas of northern China, such as Henan-Opera (豫剧), Hebei Bang-Zi (河北梆子), Ping-Ju (评剧), and Shaanxi-Opera (秦腔). With distinct regional characteristics, different Banhu subcategories express different music styles derived from local cultural contexts. Gaohu is usually used in Guangdong Opera (粤剧) or as a soprano instrument in Chinese national orchestras. Transformed from Erhu with a higher pitch range, Gaohu is placed between the knees instead of on the top of the left thigh to make a brighter sound. Zhuihu was formed in the late Qing Dynasty; with a history of over 170 years, it is prevalent in Henan and Shandong Provinces. A special difference from other HuQin is that it has a fingerboard. Zhuihu is the main accompaniment instrument of Quju-Opera (曲剧), Shandong Qin-shu (山东琴书) and Lu-Opera (吕剧); it is particularly good at imitating people's emotional voices. Zhonghu is short for Zhongyin Erhu (Alto Erhu), a reformed Erhu with a tuning a fourth or fifth lower than Erhu. All playing techniques on Erhu are suitable for playing Zhonghu (Qiao et al., 2010; Shen, 1997).

A scarcity of annotated datasets is the main obstacle to data-driven research. Unfortunately, the only available dataset containing HuQin music (Liang et al., 2019) has a small size (10.3 minutes of excerpts) and no annotations on musical excerpts. In this paper, we present **CCOM-HuQin**, the first multimodal performance dataset of **HuQin** music, which encompasses eight representative HuQin categories played by professional players majoring in HuQin from the Central Conservatory of Music (**CCOM**). In the dataset, we highlight the rich and complex **Playing Techniques (PT/PTs)**, which constitute important aspects of distinct regional and ethnic features of the HuQin family. As the first multimodal HuQin dataset, it comprises two subsets: (1) single PT recordings that include various articulations regarding velocity, dynamics, and pitch intervals. The subset naturally covers most use cases in practical excerpts, which is suitable for studying the characteristics of techniques and instrumental timbre; (2) classical musical excerpts provided with performance scores and transcription files with ground-truth PT annotations validated by professionals. For each musical piece, we provide high-quality audio and aligned videos with multiple-camera views. The dataset is publicly available on Zenodo and is provided with detailed documentation regarding the contents, annotations and examples. Anyone interested in using this dataset for research purposes will be able to access it through a request.[1]

One key challenge in creating the dataset is ground-truth transcription and annotation. We used Tony software (Mauch et al., 2015) for automatic note detection and pitch estimation, which does not perform well on HuQin music played with complex techniques. Over half of the notes required manual corrections, which is time-consuming and laborious. Another challenge is the PT annotation on the note-level transcriptions. The borders are difficult to determine for expressive articulations that naturally exist in HuQin performance. We propose annotation principles for all HuQin PT classes, hoping that the methodology will contribute to future efforts of performance dataset annotation.

Playing technique recognition systems have been developed for various musical instruments (Lostanlen et al., 2018; Wang et al., 2020) and specifically for bowed string instruments based on raw audio (Su et al., 2014;

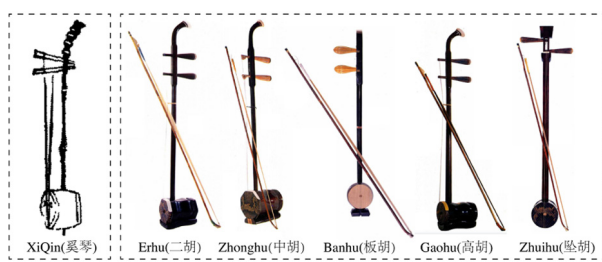

**Figure 1** Illustration of Chinese bowed instruments. The image XiQin(奚琴) is from an ancient book written by Chen (1101) and other HuQin instruments are from Liu (1992).

Ducher and Esling, 2019; Kruger and Jacobs, 2020) and other modalities (Dalmazzo and Ramírez, 2019). Beyond being a source for PT analysis, the annotated HuQin corpus serves as a challenging scenario for various fundamental MIR tasks, including pitch detection (Drugman et al., 2018), automatic transcription (Benetos et al., 2018) and score-audio alignment (Thomas et al., 2012) for music performance with abundant expressive techniques. Furthermore, the CCOM-HuQin dataset supports novel research directions for multimodal analysis (Simonetta et al., 2019) and performance-video generation (Zhu et al., 2021).

The paper is organized as follows. In Section 2, we review existing music performance datasets that have bowed-string instruments. In Section 3, we introduce a PT taxonomy for HuQin instruments. In Section 4, we describe the dataset creation procedure and elaborate on the dataset content. In Section 5, we describe the annotation process and especially highlight the annotation principles for PTs. In Section 6, we analyze the statistics of our dataset, including the count, pitch, and duration distribution from perspectives of PTs and HuQin categories. We further visualize the comparison between the score representation and F0 transcriptions to show the pitch variation of PTs. In Section 7, we perform preliminary experiments on audio and video to illustrate the potential MIR tasks that CCOM-HuQin supports. In Section 8, we conclude the paper and discuss the future prospects of the dataset.

## 2. RELATED DATASETS

In this section, we review existing music performance datasets that include bowed-string instruments. Among these, we elaborate datasets that specify PT taxonomy and annotation, and summarize their contributions to musical performance research regarding PTs. Details of these datasets are shown in Table 1.

Some large-scale datasets are built as sound banks of notes, yet missing annotation on pitch, onsets and metadata of musical excerpts. Studio On Line (SOL)[2] collects 12,000 isolated PT samples played on four bowed-string instruments. The Real World Computing Music Database (RWC) (Goto et al., 2002, 2003) contains both single PT clips and solo excerpts played on bowed-string instruments. SOL and the subset of RWC are utilized by Lostanlen et al. (2018) for instrument recognition and PT classification. Ducher and Esling (2019) publish a set of instrument PTs of cello for real-time recognition from sound banks (IPT-cello). They propose a hierarchical taxonomy to unify the PTs, which are not entirely adaptable to other ethnic instruments such as HuQin. The above datasets consider the variety of PTs, players and instruments while paying less attention to the annotation of musical content. Therefore, their usage is restricted in other fundamental MIR tasks, such as pitch analysis and transcription.

Many music performance datasets have detailed annotations despite their relatively small size, some of which include multiple modalities for performance analysis. von Coler (2018) builds a high-quality violin sound library (TU-NOTE) with annotations of instrument frequency range, note transitions and vibrato. For Indian music, Subramani and Rao (2020) build a Carnatic Violin dataset (CVD) for generative synthesis, featuring a special technique Gamaka. Elowsson and Lartillot (2021) present a Norwegian Hardanger Fiddle Dataset (HF1) with annotations of pitched onsets, offsets and emotional expression, which can be used for polyphonic transcription and music emotion recognition. Li et al. (2018) introduce the URMP dataset comprising 44 multi-instrument musical pieces for audio-visual analysis of musical performances, provided with musical scores and transcription files. Some other performance datasets focus on multimodal tasks, such as blind source separation (Montesinos et al., 2020) and motion feature analysis of players with different levels (D'Amato et al., 2020). Volpe et al. (2017) introduce the TELMI dataset to provide motion capture and EMG data to support instrument learning and teaching.

| DATASET | #INS | #PT CLIPS | EXCERPTS' DURATION | ANNOTATION | CONTENT |
|---|---|---|---|---|---|
| SOL | 4 | 12,000 clips, 15-class | N/A | N/A | audio |
| RWC | 4 | 236 clips, 5-class | 33.7min | N/A | audio |
| IPT-cello | 1 | 13.5h *, 18-class | N/A | PTs | audio |
| TU-NOTE | 1 | 1,005 clips, 4-class | 15.8min | note transitions, PTs | audio |
| CVD | 1 | 718 clips, 5-class | 37.3min | N/A | audio |
| HF1 | 1 | N/A | 42.6min | pitch, emotion | audio, transcription |
| URMP | 4 | N/A | 78min | pitch | audio, video, score, transcription |
| TELMI | 1 | N/A | N/A | N/A | audio, video, sensor data |
| CTIS | 8 | 1,072 clips, 11-class | 10.3min | N/A | audio |
| **CCOM-HuQin** | **8** | **11,992 clips, 12-class** | **77min** | **pitch, PTs** | **audio, video, score, transcription** |

**Table 1** Summary of relevant musical performance datasets. Note that all statistics are counted for bowed string instruments and the reported durations refer to solo excerpts. (*) The number of clips is not available in their documentation and the data comes from the randomly generated samples from a sound bank.

To the best of our knowledge, the Chinese Traditional Instrument Sound Database (CTIS) (Liang et al., 2019) is the only available dataset that includes HuQin music, covering pitch scales and several musical excerpts. However, the usage of CTIS is highly restricted by its small size and a lack of pitch annotations. The PT recognition results of Wang et al. (2019) show a much lower accuracy on real-world recordings than single PT clips due to unreliable labels. Compared to CTIS, our dataset has the following advantages: (1) more representative musical excerpts that cover different regions, genres, composing time and playing styles; (2) a larger scale and greater diversity of single PT recordings regarding dynamics, velocities and pitch intervals; (3) multiple camera views with high-quality audio; (4) ground-truth annotation of note-level and frame-level transcriptions. The above advantages make our dataset a vital source for various MIR tasks on HuQin music.

## 3. HUQIN PT TAXONOMY

PTs have been developing since HuQin was initially invented. In the 20th century, the PT taxonomy gradually formed an integrated system. Meanwhile, new techniques evolving from the interaction of existing ones are constantly innovated and utilized in new compositions. Referring to the definition in authoritative textbooks (Fu, 2007; Hao and Ma, 2004), related research articles (Zhao, 1999; Zeng, 2006), and considering actual usage in excerpts, we choose 12 most commonly used PTs, including eight bowing techniques and four fingering techniques. As shown in Table 2, we list HuQin PTs in Chinese and CH-Pinyin in the first two columns. The third column displays the explanations or the similar PTs (but not identical) in the violin family, and some differences are exemplified in Section 7.3. In the fourth column, we give the abbreviations depending on whether it is special in HuQin. For example, *Tremolo* is the only bowing technique that has a definition that is consistent with other bowed instruments so that *Tremolo* will be used for the rest of the paper. Several fingering techniques contain sub-classes separated by dashed lines. For the rest of this section, we describe each PT's definition and use cases in detail.

| CH | CH-PINYIN | IN VIOLIN FAMILY | ABBR. |
|---|---|---|---|
| *Bowing techniques* | | | |
| 颤弓 | ChanGong | Tremolo | Tremolo |
| 垫弓 | DianGong | N/A | DianG |
| 顿弓 | DunGong | Martelé | DunG |
| 断弓 | DuanGong | Detaché | DuanG |
| 跳弓 | TiaoGong | Spiccato | TiaoG |
| 抛弓 | PaoGong | Ricochet | PaoG |
| 击弓 | JiGong | N/A | JiG |
| 大击弓 | DaJiGong | N/A | DaJiG |
| *Fingering techniques* | | | |
| 揉弦 | RouXian | Vibrato | Vibrato |
| 滚揉 | GunRou | Rolling Vibrato | RVib |
| 压揉 | YaRou | Pressing Vibrato | PVib |
| 滑揉 | HuaRou | Sliding Vibrato | SVib |
| 滑音 | HuaYin | Portamento | Port |
| 上滑音 | Shang-Hua Yin | Upward Portamento | UPort |
| 下滑音 | Xia-Hua Yin | Downward Portamento | DPort |
| 上回滑音 | Shanghui HuaYin | Up-Down Portamento | UDPort |
| 下回滑音 | Xiahui HuaYin | Down-Up Portamento | DUPort |
| 垫指滑音 | Dianzhi HuaYin | Intermediate Portamento | IPort |
| 颤音 | ChanYin | Trill | Trill |
| 打音 | DaYin | N/A | DaYin |
| 短颤音 | DuanChanYin | Short Trill | ShTrill |
| 长颤音 | ChangChanYin | Long Trill | LoTrill |
| 拨弦 | BoXian | Pizzicato | Pizz |

**Table 2** PTs in Chinese, Pinyin, similar techniques used in the violin family if applicable and the abbreviations used in this paper. N/A means no corresponding technique in the violin family.

### 3.1 BOWING TECHNIQUES

Bowing techniques can be quantitatively parameterized by bow pressure, speed, and direction. Amplitude analysis is used to visualize bowing techniques by Root Mean Square (RMS) envelope calculation as shown in Figure 2(a-h). Two exceptions are *JiG*, and *DaJiG*, rarely used in certain pieces to simulate specific sound effects with ambiguous definitions.

***Tremolo:*** quickly pulls and pushes the bow alternately to generate continuous amplitude peaks one by one with the first bow accented in its envelope (Figure 2a). In Banhu, *Tremolo* usually expresses intense emotions in a cadenza of Hebei Bang-zi.

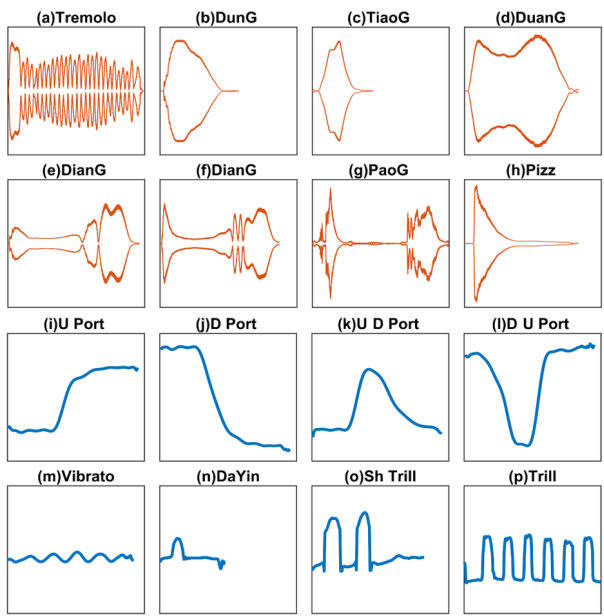

**Figure 2** RMS envelopes of bowing techniques (a-g) and a special fingering technique *Pizz* (h) with amplitude as y-axis; Pitch trajectories (i-p) of the other fingering techniques with F0 as y-axis.

**DunG:** holds the bow against the string with pressure and then releases it explosively to produce a sharp, biting attack (Figure 2b). *DunG* has a relatively more accented attack compared with *TiaoG* and *DuanG*. In Gaohu music *Yu-da Ba-jiao* (雨打芭蕉), the player uses *DunG* to imitate the crisp sound of rain beating on the plantain leaves.

**TiaoG:** controls the bow to let it bounce lightly upon the string to perform the phrase in brisk style (Figure 2c).

**DuanG:** maintains constant pressure against the string to produce a steady sound (Figure 2d).

**DianG:** quickly swings one's right wrist to produce two (Figure 2e) or three (Figure 2f) successively short and light notes after a dotted note. With fast bow speed and separate bow strokes, *DianG* is used for cheerfulness and passion in traditional Chinese opera and Telling and Singing (说唱) pieces. This technique is usually applied in Mongol-style music played on Erhu and Hebei Bang-Zi (Zeng, 2006).

**PaoG** consists of two steps: (1) perform a *TiaoG*, and lift the bow by spinning the right wrist after the sound is produced; (2) lower the bow down, then press the bow hair with the right middle and ring fingers, simultaneously producing the sound like a hoof stamp by rapidly moving the bow in the opposite direction to step (1). *PaoG* usually occurs on 16th notes in succession, expressing cheerfulness and joy and simulating a galloping horse. Figure 2g shows two separate amplitude peaks with minor noise caused by the bow stick hitting the sound box (琴筒).

## 3.2 FINGERING TECHNIQUES

As to fingering techniques, the pitch variation is the most significant factor in distinguishing one from another. We extract fundamental frequencies to visualize those differences in Figure 2(i-p). Note that *Pizz* is typically categorized as a fingering technique for the reason that players use fingers to pluck the string to make the sound, which is more suitable for amplitude analysis as shown in Figure 2h).

**Port** means sliding a finger on a string from one note to another without a discrete pitch shift. It requires the players to build a natural transition in order to make some mellow sounds. *Port* with a small pitch-interval plays an ornamenting role in phrases, lightly expressing euphemistic emotion, while *Port* with a large pitch-interval symbolizes the brave and generous spirit of people, typically in Northern China. *Port* has four sub-categories as shown in Figure 2(i-l). (1) *UPort*: slides from a lower pitch to a higher pitch. (2) *DPort*: slides from a higher pitch to a lower pitch. (3) *UDPort*: slides from the current pitch to a higher pitch and then returns to the former one. (4) *DUPort*: slides from the current pitch to a lower pitch and then returns to the former one. (5) *IPort*: slides using two or three fingers to produce a more expressive sound effect, played simultaneously with *UPort* and *DPort*.

**Vibrato** features a regular and periodic pitch change, typically characterized in terms of extent and rate. Similar to other bowed instruments, *Vibrato* is usually used to simulate the human voice (see Figure 2m). (1) *RVib*: rolls around the string by bending and stretching the first finger. Slow *RVib* expresses soulfulness and tenderness, while a fast one expresses passion, joy, and extreme sorrow. (2) *PVib*: presses the string with changing force driven by swinging the hand up and down. *PVib* usually expresses weeping accents in traditional opera arts. (3) *SVib*: slides the string back and forth for pitch variation, which is often applied for melodious, elegant, and jocular emotions in musical pieces in the cultural context of Henan, Hebei, and Shandong Province.

**Trill** consists of a rapid alternation between two notes by holding the finger on one note and resiliently hitting another note for one or several times. *Trill* can be divided into three sub-classes as shown in Figure 2(n-p): (1) *DaYin* means hitting another note once, producing an ornament for the held note; (2) *ShTrill* is for hitting another note a couple of times, typically in one second, and (3) *LoTrill* is a trill that lasts for several seconds. *Trills* with different pitch intervals represent different emotions and regional characteristics. For example, a major-second *Trill* creates a bright and magnificent sound effect; a major-third or minor-third *Trill* always appears in Mongolian Urtiin Duu (蒙古族长调).

As mentioned previously, **Pizz** is a special fingering technique with no pitch variations. Players pluck the strings with fingers inward or outward. As shown in Figure 2h, it achieves the amplitude peak instantly and releases at a slowing descent rate.

The PT taxonomy is based on the shared similarity in the HuQin family. Each HuQin category may have its unique techniques. For example, Gaohu has multiple *Appoggiatura* (回转倚音), a particular short *Trill*, as ornaments such as imitation of birds' twitter. *DaJiG* played on Erhu is usually used to mimic a galloping horse. The peculiar fingerboard of Zhuihu enables the special *Port* across multiple octaves. These samples are also included in the dataset.

## 4. CREATION OF THE DATASET

In this section, we describe the dataset creation procedure, including participating musicians, musical content selection, recording procedure, and data processing.

### 4.1 PARTICIPANTS

The eight professional players involved are all graduates majoring in HuQin performing arts with more than 15 years of experience. Some have been active performers in top Chinese folk orchestras for years. The recordist is a graduate majoring in recording arts with extensive experience in recording various concerts. Besides, we consulted professors in related fields on musical content selection, recording equipment and arrangements in the studio.

### 4.2 CONTENT SELECTION

#### 4.2.1 Single PT clips

CCOM-HuQin has 12 classes of PTs, some of which include up to four sub-classes. As described in Section 3, bowing techniques differ in bow pressure, bow speed, and bow strokes; additionally fingering techniques have different pitch intervals. Therefore we set different levels of speed (slow, medium, fast, ritardando, accelerando), dynamics (forte, mezzo forte, piano, crescendo, diminuendo), and pitch intervals (second, third, fourth) in recording single PT clips to increase the diversity of the samples and comprehensively cover cases in real-world excerpts. All settings of expressive parameters were validated and approved by participating players. To specify the parameter settings of each clip, we developed a hierarchical naming convention for PT clip labels, which is suitable for accessing and parsing in MIR tasks. Examples are detailed in the documentation provided with the dataset.

#### 4.2.2 Musical excerpts

We selected 57 musical pieces from authorized textbooks of Chinese folk instrumental music (Liu et al., 2012). These pieces represent folk music from ancient times to the present in different regions of China, showing diversity in genres, composers, composing years, and playing styles. The duration of each self-contained excerpt ranges from 30 seconds to 4 minutes. We also provide detailed metadata with the dataset such as region, date, performer, composer and description.

### 4.3 RECORDING PROCEDURE

Figure 3 displays the setting of the recording studio. We used two DPA 4011 microphones, one Neve 1073 DPA microphone amplifier and ProTools[3] as the software for audio recording. Mic 1 was positioned as the main recording equipment for the optimal sound quality, while Mic 2 was supplementary. We used three Sony X9 cameras for video recording, with 1080P resolution and built-in microphones. Camera 1 covered the front panorama of the overall movements of both hands and the body. Cameras 2 and 3 were placed for close-up shooting of the left and right hands. We made small adjustments in the camera positions before recording each performer to fully capture their body motions.

Before recording each sample, we wrote down the parameter descriptions on the whiteboard to keep the synchronization of the recordist and the player. When ready, the recordist started audio recording and the three cameras were turned on one by one, since they were controlled independently. Then we gave a simple narration for the expressive parameters of the clip to be recorded. Next, someone hit a clapperboard in the common view of three cameras and walked out of the scene. Then the player started playing, and when it ended, the clapperboard was hit again. The usage of the clapperboard remarkably facilitated audio and video synchronization in the data processing session.

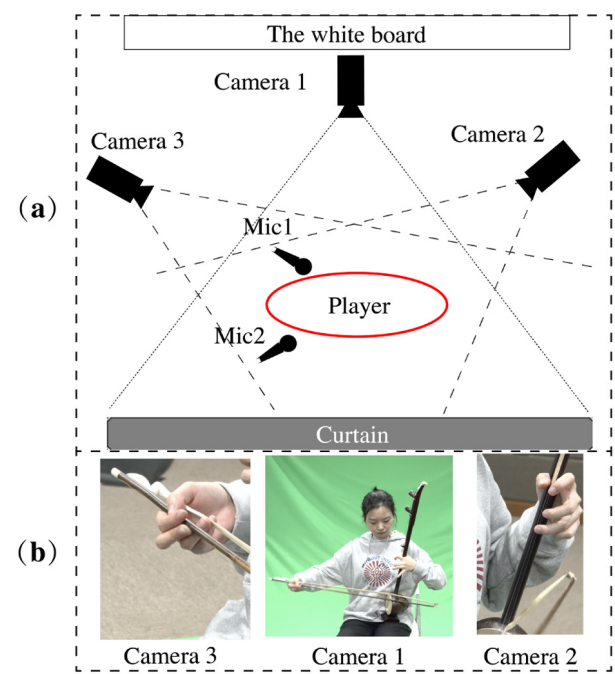

**Figure 3** (a) The floorplan of the recording studio. (b) Examples of three camera views.

## 4.4 AUDIO-VIDEO DATA PROCESSING

The CCOM-HuQin dataset contains 4.3h of audio-visual recordings. Audio is in WAV format with a sampling rate of 48 kHz and a 32-bit float depth. Videos are in MP4 format encoded with H.264, 1080P resolution (1920 × 1080), and 29.97 fps. We utilized Adobe Premiere Pro[4] for audio-video processing. Firstly, for each sample in the dataset, we replaced the original audio in the video recordings with the high-quality audio recorded by DPA 4011 microphones. Then we aligned the three videos by the shared audio track using *Merge-Clips* function. Next, since one recording sample contains dozens of PT single clips, we split the sample into single clips manually, with each clip containing the complete note onset and offset. Sometimes several clips were played consecutively so that onsets and offsets may overlap. We prioritized the next clip's onsets over the previous clip's offsets, considering onsets are more important to distinguish PTs. Finally, we exported the HQ audio and three aligned videos.

## 5. GROUND-TRUTH ANNOTATION

Beyond recordings, each musical excerpt is provided with (1) a graphical score in PDF format and symbolic encodings in MusicXML format, both with PT annotations; (2) note-level transcriptions with PT annotations and frame-level pitch trajectories in CSV format. In this section, we introduce how we generate scores, note-level transcriptions and frame-level pitch trajectories with a set of ground-truth annotation rules. The entire annotation pipeline is shown in Figure 4.

### 5.1 SCORE ANNOTATION

To start with, players were asked to annotate the applied techniques for all the notes on the performance sheet, referring to the audio and video recordings. Note that the symbols may vary with different players for the performance notations, although most follow the same convention. Therefore, we verified the annotation reference table with players after they finished.

After players gave feedback on their performance sheets with hand-written PT symbols, we invited professional score producers, mostly majoring in HuQin performing arts, to manually type the scores using MuseScore 3.0,[5] including all the PT symbols. Obvious handwriting errors in the original sheets were corrected during this process based on the recordings.

### 5.2 AUDIO ANNOTATION
#### 5.2.1 Transcription

For transcription, we used the Tony software (Mauch et al., 2015) to extract note-level sequences and frame-level pitch tracks from the monophonic recordings. For note sequences, we performed manual corrections, including note adding, splitting and merging. We observed that bowing techniques with short durations, such as *TiaoG* and *DunG*, are often missing in the automatic analysis results, so manual note adding was required. For pitch trajectories, notes with fingering techniques that involve large pitch intervals required most adjustments, such as *Port* and *Trill*, where missing parts needed to be completed. Over half of the notes required manual corrections. On average, it took about 30 minutes to finish the corrections on a 30-second excerpt. It suggests that automatic note detection and pitch estimation are challenging tasks for HuQin with expressive performance techniques.

#### 5.2.2 PT annotation

Along with transcriptions, we annotated the note sequences with PTs using the annotated score. The primary difficulty was identifying the boundaries of techniques played without separation. Additionally, since one PT could be performed on multiple notes in the recordings, it was necessary to label Begin (B), Intermediate (I) and End (E) to indicate the sequence of the applied technique, which is named as *BIE* annotation pattern, a typical method utilized for sequence labeling in Named Entity Recognition (Konkol and Konopík, 2015). According to the natural characteristics of certain techniques and their practical use cases in excerpts, the annotation principles were developed as follows.

Figure 5(a-c) illustrates the annotation principles for bowing techniques that cover multiple notes. *Tremolo* features rapid bow motion and generates dozens of notes. Thus the beginning note was annotated as *Tremolo-B*, followed by a number of *Tremolo-I*s and the end note as *Tremolo-E*. *DianG* begins with a dotted note followed by two or three short notes, and *PaoG* usually involves three short notes. They both follow the *BIE* pattern. Other bowing techniques were annotated on a single note, which was straightforward, and we decided not to show them for simplicity.

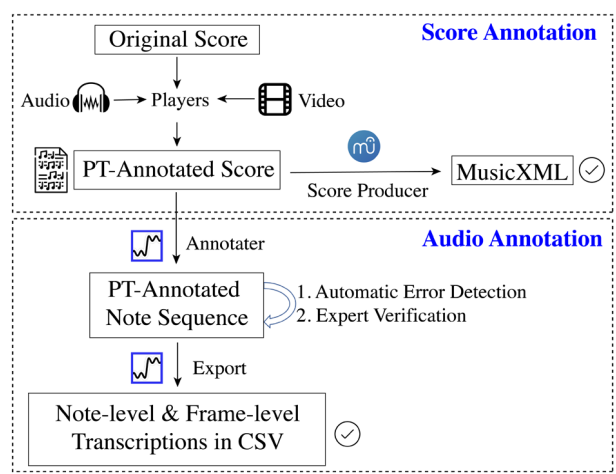

**Figure 4** The annotation pipeline.

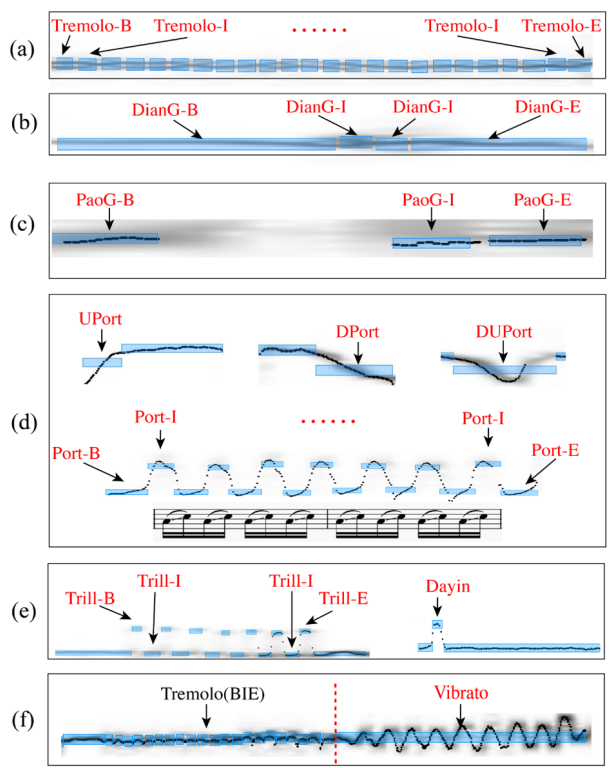

**Figure 5** PT annotation examples of (a) *Tremolo*; (b) *DianG*; (c) *PaoG*; (d) *Port*; (e) *Trill*; (f) *Vibrato*.

*Port* has the most variations among all PT classes in musical excerpts. The main difficulty is to determine boundaries of *Port* that involve uncertain sliding motion. We show examples of *Port* subclasses in Figure 5d. Yang (2016) proposed to annotate a *Port* from the midpoint of the beginning note to the midpoint of the target note. However, one problem with following this rule is that the automatically-detected note needs to be split into multiple parts. This manual correction conflicts with the note segmentation for the purpose of transcriptions. Besides, multiple techniques such as *Port* and *Vibrato* could be played simultaneously or continuously on a note. Annotation for *Port* should not break other PT annotation rules. Based on the guidelines for the labeling of *Port* transitions introduced by von Coler and Lerch (2014), and considering the complex cases existing in HuQin performance, we annotated *Port* according to the three rules: (1) Only the transition segment of the *Port* is annotated. The top line (*UPort*, *DPort* and *DUPort*) in Figure 5d shows the examples applicable to this rule; (2) In rare cases, consecutive multiple *Port*s are played between a series of notes without a clear boundary. They are usually notated with slurs in musical scores. We find it hard to confine the transition part, so the BIE pattern indicates the inter-connected *Port* notes as a whole. The bottom line in Figure 5d exemplifies this situation; (3) Subject to (1) and (2), we try to avoid manual splitting of the automatically detected notes.

Figure 5e demonstrates a typical *Trill* that consists of a sequence of notes and a particular case of the sub-class *DaYin*, which is unique in HuQin music. *DaYin* means hitting another note for one time very rapidly (around 0.1s), which is usually missing in automatic detection. New notes were created and annotated.

*Vibrato* can be played for the entire or just part of a long note. Figure 5f shows an example of late vibrato, where a *Tremolo* is played before the *Vibrato*. Therefore, we split the notes and annotated where *Vibrato* is played.

### 5.2.3 Annotation verification

In this section, we elaborate on the verification process for refinement and accuracy. Annotation of audio was based on the score annotations that were completed by the musicians and manually validated by the professional score producers. Thus, the verification process of audio annotation guarantees the alignment of score annotation and recording annotation.

Three professionals knowledgeable about HuQin performance participated in the annotation process of the dataset. After the initial annotation, spelling and semantic errors violating *BIE* syntax in annotated PTs were automatically detected. Then, two professionals checked the annotation file in exchange for correcting the errors against the annotation principles and completing the missing labels based on the annotated musical score. After that, the third professional decided whether to adopt the correction or not. Therefore, the final annotation was verified by at least three professionals. On average, the verification process took about five minutes for a 30-second excerpt, which was efficient compared to 30 minutes for the initial annotation work.

As soon as the transcription, annotation, and verification processes were complete, we exported the note-level transcription with PT annotations and the corresponding frame-level pitch track as CSV files. In summary, annotation on excerpts is challenging, time-consuming, and requires music expertise in HuQin. These guidelines are effective based on the coordination between professional players and other participants. Hopefully, our endeavors will shed light on future efforts of music dataset creation.

## 6. STATISTICS AND ANALYSIS

In this section, we present the dataset statistics. Firstly, the count and duration distribution of single PT clips and the pitch distribution of different HuQin categories are given. Secondly, we show the percentages of annotated notes in all musical excerpts and compare them among PT classes and instruments. Furthermore, we visualize the ground-truth F0 trajectories and score representation of representative samples to demonstrate the characteristics of pitch-variant techniques.

## 6.1 SINGLE PT CLIPS

The dataset contains single clips of 12-class PTs played in eight HuQin categories, with a total duration of 2h 52min. As described in Section 3, one special PT, *DaJiG*, is to mimic the sound effect of a galloping horse and has no regular characteristics of pitch or duration. Thus in this section, we demonstrate the statistics of the other 11 PT classes.

### 6.1.1 Count distribution

Figure 6a shows the PT clip count by each HuQin category sorted in descending order, where Erhu and Banhu (including four subcategories) are the most common. Figure 6b shows the count distribution of PT clips, which is similar to the practical use cases in excerpts. *Port* and *Vibrato* have the highest proportion of all PTs because of their frequent usage and diversities in subclasses. *JiG* has the minimum number since it is rarely used in practice.

### 6.1.2 Pitch distribution

To obtain the pitch distribution of HuQin categories in the dataset, we used the pYin (Mauch and Dixon, 2014) function in Librosa (McFee et al., 2015) to estimate the fundamental frequencies at a frame hop length of 23ms for all single PT clips. Figure 6c shows the pitch distribution in ascending order for eight categories of HuQin. The average pitch ranges from lower C5 for Zhonghu to higher C6 for SBanhu. The shades represent one standard deviation area. This reflects the common pitch ranges of HuQin instruments.

### 6.1.3 Duration distribution

Figure 6d shows the duration distribution of single PT clips. For each class, the mean duration ranges from 0.1s to 2s. *TiaoG* and *DunG* have a shorter average duration since they are played by dropping the bow on the string and lifted immediately. *Vibrato* and *Tremolo* have a longer duration and often are used on a sustained note produced by periodic hand motion. *Trills* have the largest standard deviations. The subclasses of *Trills* include *DaYin* (two or three notes in 0.5s), short *Trills* (five notes in 1s) and long *Trills* (ten or more notes in 3-5s). *Vibratos* also show apparent duration fluctuations for the various playing rates and the number of periods. The duration distribution implies the diversity of recordings regarding the various expressive parameter settings in CCOM-HuQin.

## 6.2 EXCERPTS

### 6.2.1 Annotation count

Figure 7a shows the number of annotated notes in excerpts sorted in descending order in terms of HuQin instruments and PTs. Erhu excerpts have the largest note count and achieve the highest percentage of annotation. On average, the percentage of annotated notes in all excerpts reaches 43%. As shown in Figure 7b, *Port*, *Vibrato* and *Trill* are the top-3 most applied PTs in HuQin excerpts for melodic expression.

### 6.2.2 Pitch track visualization

One key challenge in HuQin music transcription is complex pitch-varying techniques. As mentioned in Section 5.2, ground-truth F0 annotation demands time-consuming and expertise-involved work based on the automatically analyzed results. Here we show specific examples of pitch track differences between score representations and audio recordings. For this purpose, we extracted the pitch sequence from the score and temporally aligned it with note-level audio transcriptions using the onset and duration labels. Then we visualize the aligned score representation and frame-level pitch trajectory with PT

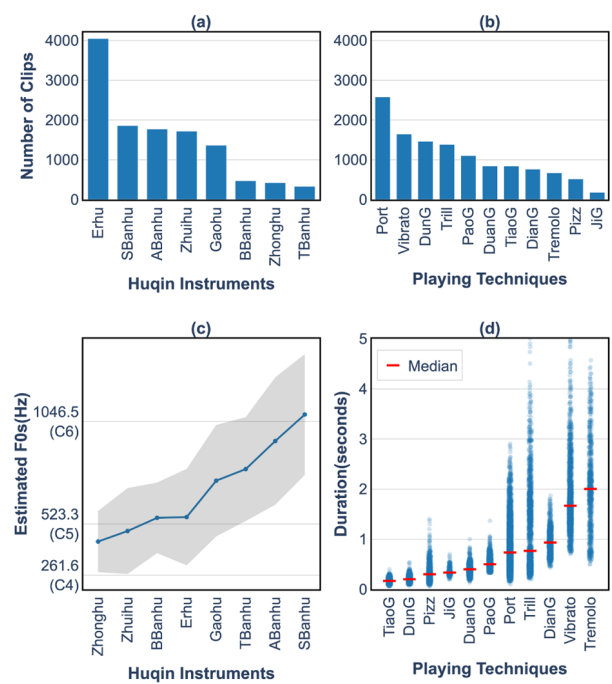

**Figure 6** Statistics for PT short clips: (a) count distribution for HuQin instruments; (b) count distribution of PTs; (c) pitch distribution of HuQin instruments (A4=440Hz); (d) duration distribution of PTs.

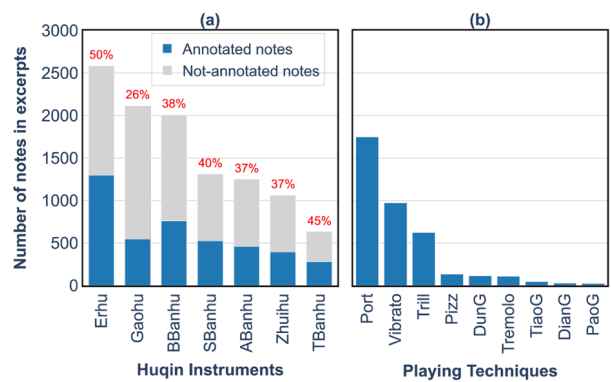

**Figure 7** Number of notes in excerpts for (a) HuQin instruments, with the percentage of annotated notes; (b) PT distribution of all annotations.

annotation using the time-frequency as x-y axes, as shown in Figure 8.

Banhu performance has abundant *Ports* and *Vibratos*. *UPort* and *DPort* are typically applied at the beginning or end of a note. As shown in Figure 8a, the *UPort* is sliding from a lower pitch to the main pitch at the beginning. *DUPort* is played at the mid-point of a long note to emphasize the original note for strong emotional feelings. The *PVib* has a larger playing extent than other vibratos, typically used in Shanxi local songs to express passion. Figure 8b presents an example of Gaohu music. It shows a particular *ShTrill*, which involves three notes to produce a decoration, while the *Trill* in Figure 8a is a longer one with alternation between two notes. The *RVib*, a commonly used *Vibrato* by Gaohu, is usually intended for a graceful and beautiful melody. Figure 8c shows one excerpt played on Erhu, including a distinctive *Port* with pressing. This *Port* imitates human crying in a mimic way. The pitch track shows a relatively large extent and high rate for pressing motion for a realistic weeping sound effect. *DaYin* shown in Figure 8a and c represent its frequent usage at the beginning of a note for decoration. Figure 8d exemplifies rich expressiveness in Zhuihu performance. Several large pitch-interval *Ports* indicate the emotional feelings and aesthetics of local opera arts. The special usage of *DPort* shows a vivid imitation of human laughter, often performed in operas. Note that the score representation for this *DPort* is as simple as repeated notes of the same pitch. Performers need to apply special techniques to achieve the real effects (Liu, 1986; Li, 2007).

To sum up, the pitch sequence in a musical score is far from capable of representing variation in performance. Moreover, the articulation implicated by each PT depends to a large extent on the individual understanding, cultural background and performance style of HuQin players. Concretely, these are embodied in differences in bowing parameters, pitch variation patterns, and other implicit aspects to be explored. In other words, the traditional music notation schema and transcription methods are not completely applicable to HuQin music, which is the basic motivation of our work on constructing the dataset and proposing a set of novel transcription and annotation paradigms.

## 7. POTENTIAL APPLICATIONS

CCOM-HuQin has great potential in MIR research. For music performance analysis, audio and video can be jointly learned to recognize PTs or to model the performer's style and music genre. Detailed annotations enable setting benchmarks on traditional MIR tasks such as onset detection, pitch estimation, and audio-score alignment on HuQin music. Moreover, high-quality audio-visual recordings with comprehensive labels may contribute to audio-visual generation on global body movements and local hand poses. In this section, we show two preliminary PT classification experiments and present some findings on the audio and video of single PT clips. We further compare the difference in PTs used in Erhu and the violin for a case study of cross-cultural music analysis. Source code is released at Github.[6]

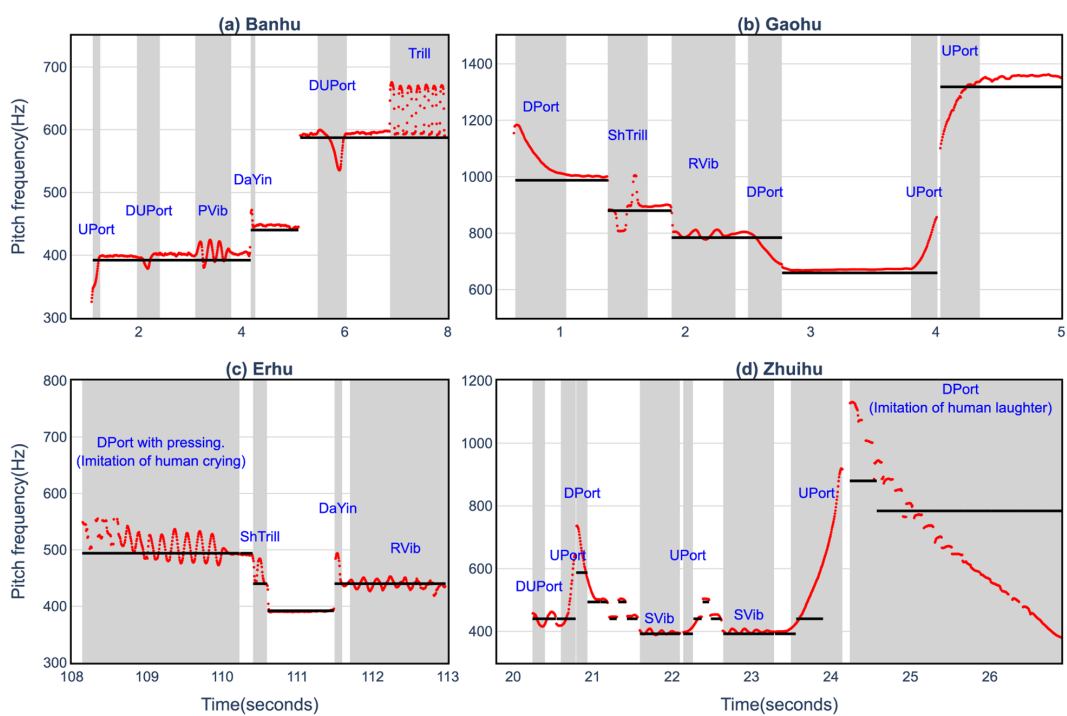

**Figure 8** Pitch variation visualization of ground-truth pitch tracks (red) and score representation (black) for typical excerpts played on (a) Banhu, (b) Gaohu, (c) Erhu, (d) Zhuihu.

## 7.1 AUTOMATIC PT CLASSIFICATION ON AUDIO

We perform a baseline PT classification experiment on the single PT clips to evaluate the quality of CCOM-HuQin, especially for diversity and generalization. We compare the performance of models on homogeneous and heterogeneous corpora, like Ducher and Esling (2019). Homogeneous corpus means the testing set has the same source as the training set, while heterogeneous corpus means they come from different sources. We also include the CTIS dataset (Liang et al., 2019) for comparison. Using our proposed taxonomy, we obtain nine classes of PTs for training and testing sets, including *Tremolo*, *Trill*, *DianG*, *DunG*, *PaoG*, *Vibrato*, *Port* and *Pizz*. They are the most common PTs used in real-world musical excerpts and could be played for all HuQin subcategories. The statistics of the number of notes in the corpus are shown in Table 3. The mel-spectrogram is used as the input feature. We used CNN (Zhang et al., 2016) and CRNN (Choi et al., 2017) as classifiers and Adam Optimizer (Kingma and Ba, 2015) for stochastic gradient descent optimization. The homogeneous and heterogeneous settings of train/validation/test sets and overall results are shown in Table 4.

Results on homogeneous datasets show that the model performs slightly better on CTIS-I than on CCOM-HuQin, due to the fact that CCOM-HuQin has a larger variety of instruments, PTs and expressive parameters than CTIS-I. For heterogeneous results, the F1 score using CCOM-HuQin as training data is significantly higher than using CTIS, suggesting that CCOM-HuQin is more comprehensive and thus suitable for machine-learning algorithms. Besides, CRNN outperforms CNN in all dataset settings. The classes that are easily confused by the CNN have a relatively short duration of around one second, whereas these errors are avoided with the CRNN. That illustrates the combination of the recurrent time sequence model can better learn the difference between PTs, especially for those with shorter durations.

On the other hand, some errors are shared in CNN and CRNN configurations. Figure 9a shows that *DaYin* is likely to be mistaken as *Port* and *Trill* as *Vibrato*. Figure 9b visualizes those confusing pairs using spectrograms. Specifically, the spectrogram shape of *DaYin* has a similar shape with *DUPort*, which is embodied as a transition to a lower pitch and then returning to the original pitch. *Trill* has a sinusoidal shape, similar to *Vibrato*. In the following section, we used video samples to identify the differences between these two sets of easily confused PTs.

## 7.2 HAND POSE DETECTION ON VIDEO

Videos showing hand pose provide supplementary information for PT recognition (Li et al., 2017). We extract three-axis data points of 21 left-hand landmarks for the two pairs of fingering techniques using the hand pose estimation module in MediaPipe (Zhang et al., 2020), which returns a three-axis coordinate for each hand landmark. The z-axis represents relative depth in a 2D frame. Based on the extracted landmarks on each axis between frames of samples, we conduct supervised binary classification experiments respectively between *DaYin* and *Port*, *Trill* and *Vibrato* samples using a support vector machine

| DATASET | BOWED-STRING INSTRUMENTS | COUNT |
|---|---|---|
| CTIS-I | Erhu | 787 |
| CTIS-II | Banhu, Soprano Banhu, Alto Banhu, XiQin, Zhonghu, Zhuihu | 285 |
| Hybrid | Erhu | 252 |
| **CCOM-HuQin** | **Erhu, Soprano Banhu, Alto Banhu, Tenor Banhu, Bass Banhu, Gaohu, Zhonghu, Zhuihu** | 11,014 |

**Table 3** Statistics of training and testing sets.

| DATASET | CNN | CRNN |
|---|---|---|
| *Homogeneous* | | |
| CTIS-I | 97.54% | 99.19% |
| **CCOM-HuQin** | 96.07% | 97.85% |
| *Heterogeneous (Train Validation / Test)* | | |
| CTIS-I / CTIS-II + Hybrid | 69.21% | 70.39% |
| **CCOM-HuQin** / CTIS-I & II + Hybrid | 77.01% | **87.01%** |

**Table 4** F1 score of classification results.

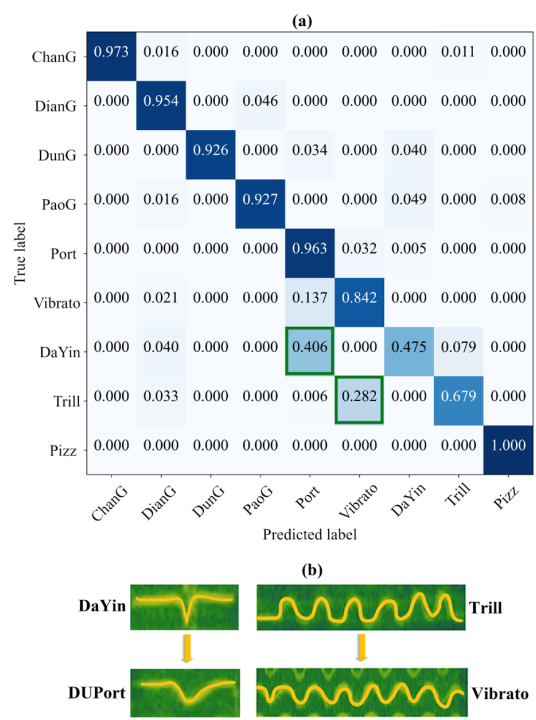

**Figure 9** (a) Nine-class confusion matrix of CRNN classification result. (b) Spectrogram examples of two pairs of easily confused PTs.

(SVM) classifier. The average accuracy of all 21 landmarks is shown in Table 5. The algorithm achieves the highest binary classification accuracy on the x-axis, 87.24% between *DaYin* and *Port*, 77.82% between *Trill* and *Vibrato*.

Then we select representative samples of *Vibrato* and *Trill* and visualize the hand key-point's change in video frames on x-, y- and z-axes, as shown in Figure 10. We choose the fingertip that hits the strings as the key point (index finger for *Vibrato* and middle finger for *Trill*). In Figure 10a, periodic change is observed in both techniques. *Vibrato* has a gentle starting phase, while the *Trill* begins with a sudden upward motion, which is the most evident on the x-axis. The starred points correspond to Figure 10b, where we highlight the key points in video frames. The complementary information provided by video may improve the performance of PT classification results.

## 7.3 PT COMPARISON BETWEEN ERHU AND VIOLIN

In this section, we choose two representative examples respectively from bowing and fingering techniques to compare PTs used for Erhu and the violin. As mentioned in Section 3, there are some similar but not identical bowing techniques in the violin family. We select *PaoG/Ricochet* as an example to explain their difference in playing methods and cultural meaning. As shown in Figure 11, Erhu's *PaoG* contains *TiaoG* (decribed in Section 3), and usually appears with two or three peaks of the envelope, imitating galloping horses. The violin's ricochet consists of throwing the bow onto the string using the upper third of the bow in a downward direction so that it bounces and produces a series of rapid notes (Tsou, 2001). It often appears as light *staccato* and was applied by nineteenth-century virtuosi like Paganini.

For fingering techniques, we get statistics of *Port* in Erhu excerpts from CCOM-HuQin, which covers a variety of compositions as described in Section 4. For violin, we obtain 16.2 minutes of solo excerpts from two publicly available datasets (von Coler, 2018; Thickstun et al., 2016) and our own recordings. Violin excerpts are Western classical music pieces of nine composers from the Baroque period to the 20th century. Different playing styles are included, since they are from multiple data sources. We manually labeled *Port* based on both auditory perception and visual spectrograms. As shown in Table 6, Erhu excerpts have approximately five times as many *Ports* as the violin excerpts within the same length of time, which implies that *Port* constitutes an important aspect of Chinese instrumental music. Further studies on Port parameters, such as dynamics, slopes and pitch intervals can be performed for more interesting findings on comparisons between HuQin and violin music.

| COORDINATES | *DAYIN/PORT* | *TRILL/VIBRATO* |
|---|---|---|
| **X-axis** | **87.24%** | **77.82%** |
| Y-axis | 86.40% | 72.30% |
| Z-axis | 86.31% | 76.33% |

**Table 5** SVM classification accuracy on two pairs of confusing PTs.

## 8. CONCLUSION AND FUTURE WORK

In this paper, we presented CCOM-HuQin, the first annotated multimodal performance dataset of HuQin music. With abundant PTs, this dataset is a challenging source for many MIR tasks in performance analysis and synthesis. We described the HuQin PT taxonomy and its important roles in HuQin musical expression. We developed a pipeline that enables coordination between players, recordists, score producers and annotators, for an effective recording and annotation process. Furthermore, we showed the statistics of the dataset and visualized the comparison between score notation and ground-truth pitch trajectories. Finally, we demonstrated the dataset's potential applications through two preliminary experiments and a case study of PT comparison between HuQin and the violin. The main purpose of this paper is to create a standardized HuQin dataset by featuring PTs and maintaining their cultural characteristics. It is challenging but meaningful for HuQin performance research from the interdisciplinary perspectives of music and technology.

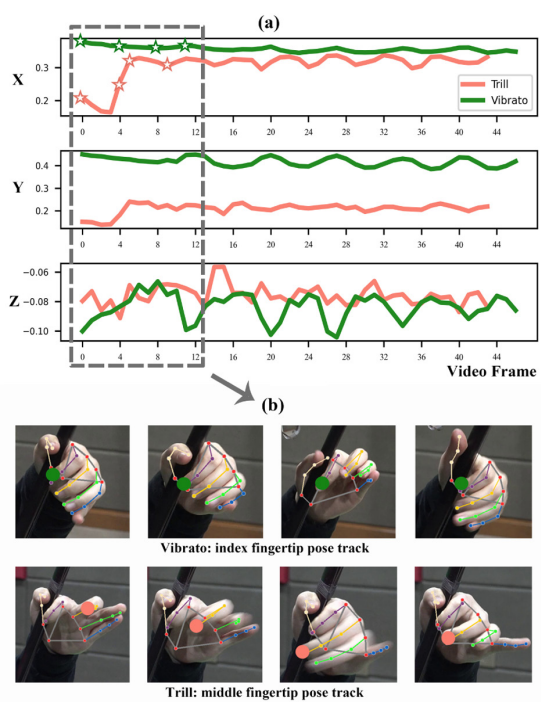

**Figure 10** Hand pose visualization of *Trill* and *Vibrato* in (a) key-point's change on x-, y- and z-axis and (b) selected video frames with fingertip labels.

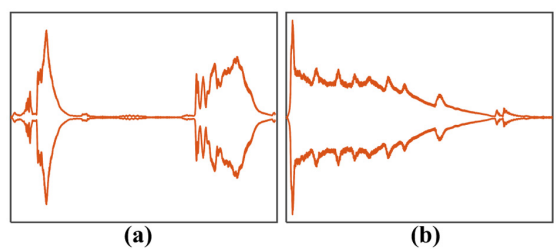

**Figure 11** Comparison of RMS envelopes between (a) *PaoG* on Erhu and (b) *ricochet* on the violin.

| EXCERPTS | DURATION | #PORT | #PORT/MIN |
|---|---|---|---|
| Erhu | 22.7min | 484 | 21.4 |
| Violin | 16.2min | 70 | 4.3 |

**Table 6** Comparison of the number of *Ports* between Erhu and violin music.

On the other hand, with aligned audios, multiple-view videos and scores, CCOM-HuQin can further be used for multimodal scenarios, such as gesture and body movement generation in music performance given the audio. In the future, the dataset will be extended with more HuQin instruments and musical excerpts, including traditional and contemporary compositions, based on the standardized procedure we proposed. Moreover, we will extend the annotation principles to cover more playing styles of HuQin music and enable the entire annotation process to be more interactive between different annotators, e.g. composers, players, teachers and even amateur learners.

### NOTES

1. https://doi.org/10.5281/zenodo.6957454.
2. https://forum.ircam.fr/projects/detail/fullsol/.
3. https://www.avid.com/pro-tools.
4. https://www.adobe.com/products/premiere.html.
5. https://musescore.org/.
6. https://github.com/zhangyubella/CCOM-HuQin.

### ETHICS AND CONSENT

All participants signed an informed consent form to be part of generating the public dataset.

### ACKNOWLEDGEMENTS

We thank Kai Chen, Li Li, Zengqi Zhang, Runze Li, Peize Li, Zhaoxing Wang, Chuting Wang for participating in performance and annotation and providing professional advice on building the dataset, Yiru Xu and Xueying Wang for their recording expertise. We also thank Prof. Zijin Li for her constructive comments that greatly improved the paper.

### FUNDING INFORMATION

This work was supported by Special Program of National Natural Science Foundation of China (Grant No. T2341003), Advanced Discipline Construction Project of Beijing Universities, Major Program of National Social Science Fund of China (Grant No. 21ZD19), Nation Culture and Tourism Technological Innovation Engineering Project (Research and Application of 3D Music) and Construction of Digital Cultural Heritage Platform for Ethnic Musical Instruments in Southwest China(Grant No.22VJXG012).

### COMPETING INTERESTS

The authors have no competing interests to declare.

### AUTHOR CONTRIBUTIONS

Yu Zhang and Ziya Zhou the authors made equal contribution to this work.

### AUTHOR AFFILIATIONS

**Yu Zhang** orcid.org/0000-0002-0310-3705
Department of AI Music and Music Information Technology, Central Conservatory of Music, Beijing, China

**Ziya Zhou** orcid.org/0000-0002-8113-9842
Department of AI Music and Music Information Technology, Central Conservatory of Music, Beijing, China

**Xiaobing Li**
Department of AI Music and Music Information Technology, Central Conservatory of Music, Beijing, China

**Feng Yu**
Department of AI Music and Music Information Technology, Central Conservatory of Music, Beijing, China

**Maosong Sun** orcid.org/0000-0002-6011-6115
Department of AI Music and Music Information Technology, Central Conservatory of Music, Beijing, China; Department of Computer Science and Technology, Tsinghua University, Beijing, China